# Pressure tuning of optical reflectivity in LuH$_2$

Xuan Zhao[1,2#], Pengfei Shan[1,2#], Ningning Wang[1,2], Yunliang Li[1,2], Yang Xu[1,2*], and Jinguang Cheng[1,2*]

[1]Beijing National Laboratory for Condensed Matter Physics and Institute of Physics, Chinese Academy of Sciences, Beijing 100190, China

[2]School of Physical Sciences, University of Chinese Academy of Sciences, Beijing 100190, China

# These authors contributed equally to this work.

*Corresponding authors: yang.xu@iphy.ac.cn; jgcheng@iphy.ac.cn

**Abstract**

Recently, the claim of room-temperature superconductivity in nitrogen-doped lutetium hydride at near-ambient conditions has attracted tremendous attention. Criticism of the work rises shortly, while further explorations are needed to settle the dispute. One of the intriguing observations is the pressured-induced color change, which has been reproduced in the lutetium dihydride LuH$_2$ while its mechanism remains unclear. Through optical reflectivity measurements of LuH$_2$ in the visible to near-infrared region, we observe strong light absorption next to the sharp plasmon resonance, which continuously shifts to higher energies with increasing pressure. It gives rise to the increased reflection of red light and suppressed reflection of blue light. Our work sheds light on resolving the puzzles regarding the pressure induced color change in LuH$_2$.

In the last few years, interest in the rare-earth hydrides has been reinvigorated owing to their potentials for hosting room-temperature superconductivity under achievable pressures. Prominent examples include LaH$_{10}$ and YH$_9$, in which the superconducting transition temperature ($T_c$) approaches 260 K at ~200 GPa [1–4]. Nitrogen doped lutetium hydride was recently claimed to exhibit possible superconductivity at a room temperature of 294 K under near-ambient pressure of ~1 GPa, concomitant with remarkable color changes from lustrous blue to pink and subsequently bright red with increasing pressures (a few GPa) [5]. Similar color changes have been observed shortly by two independent studies in nitrogen-free lutetium dihydride LuH$_2$ but no superconductivity was observed up to 7 GPa [6,7]. The peculiar color changes motivate us for further investigation using optical spectroscopy. Here we report the optical reflectivity measured for LuH$_2$ from visible to near-infrared region (420~900 nm) at different pressures up to ~14 GPa. We observe strong absorption of red light close to a sharp plasmon edge near 750 nm at ambient conditions. Such plasmon edge blueshifts with a pressure coefficient of 9.4~12.6 meV/GPa, resulting in the increase of red-light reflection and blue-light absorption. Our work thus unveils the dominant role of pressure modulated plasma frequency $\omega_p$ for the color changes observed in LuH$_2$.

We start with the commercially purchased LuH$_2$ powder (99.9%) from JiangXi Viilaa Metal Material Co., Ltd. Rietveld refinements on the powder XRD pattern (Fig. S1) confirmed that the major phase is the cubic LuH$_2$ with the fluorite structure (space group $Fm\bar{3}m$), coexisting with a minor phase of Lu metal and unidentified phases marked with asterisks. The blue shining grains of LuH$_2$ were pre-compressed to flat platelets (measuring 5-10 μm in thickness) using WC anvils. These platelets were loaded into the symmetric diamond anvil cells (DACs) with 300 μm culets and T301 stainless steel gaskets, in the middle of which ~180 μm diameter hole was drilled and served as the sample chamber. Soft KBr was employed as the pressure transmitting medium. Silver is known to possess near-unity reflectivity in the visible to near-infrared spectral range and thus was mounted inside the DAC as the reflection reference. The pressure inside DAC was determined from the shift of ruby fluorescence lines. For the optical reflectivity measurements [8,9], broadband radiation from a supercontinuum laser (NKT FIU-6) was guided and focused into the DAC with a 10× objective. The reflected beam was collected by the same objective and detected by a spectrometer coupled with a CCD camera (Princeton Instrument). The reflectivity of LuH$_2$ is obtained by calculating the ratio between the reflection intensity from LuH$_2$ and that from the silver mirror. Sample observations and snapshots are conducted without alternating the measurement system (see more details in Fig. S2).

Displayed in Fig. 1 are the micrographs of LuH$_2$ sample 1 at different pressures. The sample exhibits a shiny blue color at ambient conditions and immediately turns into violet after direct contact with the diamond culet. The color of the sample changes to dark red around ~3 GPa and then orange at pressure above ~12 GPa. These new results are supplementary to our previous work and confirmed the pressure-induced continuous color changes in LuH$_2$ [6]. The corresponding reflectivity spectra are shown in Fig. 2a. At ambient conditions, the reflectivity is relatively large at longer wavelengths above ~750 nm. Meanwhile, there is a sharp suppression of the red reflection (<~0.03) near 675 nm and the reflectivity goes up at shorter wavelengths. These are typical features of a plasmon resonance, near which the collective excitation of electrons causes sign change of the real part of the dielectric function. The reflectivity hence has a pronounced minimum near a sharp plasmon edge. It also explains the initial bluish color seen in Fig. 1 and the results are consistent with the literature [10]. With increasing pressure, the plasmon edge continuously shifts to higher energies. The reflection of blue and red at representative wavelengths (495 and 675 nm, respectively, highlighted by the dashed vertical lines in Fig. 2a) are extracted and shown in Fig. 2b. The boost of red reflection and the continuous suppression of the blue reflection confirm the color change observed through the optical microscope (Fig. 1). In Fig. 2c, we estimate the plasmon energy $\hbar\omega_p$ (or the plasmon edge) by finding the peak in the reflectivity derivatives. The extracted peak energy as a function of pressure is plotted in Fig. 2d for three different samples. The upper surface of the sample 1-2 under pressure is in direct contact with the DAC, while for sample 3, the upper surface is in contact with KBr. As can be seen, all the three samples feature similar positive pressure coefficients, i.e., 12.4±0.7, 12.6±0.8, and 9.4±0.8 meV/GPa, respectively. We also noticed discontinuity of the reflectivity (Fig. 2b) and plasmon energy (Fig. 2d) upon increasing pressure from ambient for sample 1 (similar for sample 2). It is likely due to the fact that the sample experiences a considerable stress that cannot be accurately measured by the ruby calibration during the initial compression on the sample directly with DAC. By contrast, the surface of sample 3 touches the soft KBr and should experience less stress during the initial compression. However, the strong cavity interference severely hinders the accurate measurement of the reflectivity spectrum and causes larger uncertainty of the extracted pressure coefficient (see more details in Fig. S4).

To summarize, the reflectivity of $LuH_2$ changes significantly in the visible range due the plasmon resonance, which continuously shifts to higher energies with increasing pressure. Our work sheds light on resolving the puzzles regarding the pressure modulated color change in lutetium dihydride [5–7].

**Acknowledgements**

This work is supported by the National Natural Science Foundation of China (Grant Nos. 12025408, 11921004, 11888101), the Beijing Natural Science Foundation (Z190008), the National Key R&D Program of China (2021YFA1400200), the Strategic Priority Research Program of CAS (XDB33000000).

**References**

[1] Drozdov AP, Kong PP, Minkov VS, Besedin SP, Kuzovnikov MA, Mozaffari S, et al. Superconductivity at 250 K in lanthanum hydride under high pressures. Nature 2019; 569:528–531.

[2] Somayazulu M, Ahart M, Mishra AK, Geballe ZM, Baldini M, Meng Y, et al. Evidence for Superconductivity above 260 K in Lanthanum Superhydride at Megabar Pressures. Phys Rev Lett 2019; 122:027001.

[3] Kong P, Minkov VS, Kuzovnikov MA, Drozdov AP, Besedin SP, Mozaffari S, et al. Superconductivity up to 243 K in the yttrium-hydrogen system under high pressure. Nat Commun 2021; 12:5075.

[4] Snider E, Dasenbrock-Gammon N, McBride R, Wang X, Meyers N, Lawler K V., et al. Synthesis of Yttrium Superhydride Superconductor with a Transition Temperature up to 262 K by Catalytic Hydrogenation at High Pressures. Phys Rev Lett 2021; 126:117003.

[5] Dasenbrock-Gammon N, Snider E, McBride R, Pasan H, Durkee D, Khalvashi-Sutter N, et al. Evidence of near-ambient superconductivity in a N-doped lutetium hydride. Nature 2023; 615:244–250.

[6] Shan P, Wang N, Zheng X, Qiu Q, Peng Y, Cheng J. Pressure-induced color change in the lutetium dihydride $LuH_2$. Chin Phys Lett 2023; 40:046101.

[7] Ming X, Zhang Y-J, Zhu X, Li Q, He C, Liu Y, et al. Absence of near-ambient superconductivity in $LuH_{2\pm x}N_y$. arXiv 2023; 2303.08759.

[8] Welber B. Micro-optic system for reflectance measurements at pressures to 70 kilobar. Rev Sci Instrum 1977; 48:395–398.

[9] Seagle CT, Dolan DH. Note: Visible reflectivity system for high-pressure studies. Rev Sci Instrum 2013; 84:066104.

[10] Weaver JH, Rosei R, Peterson DT. Electronic structure of metal hydrides. I. Optical studies of $ScH_2$, $YH_2$, and $LuH_2$. Phys Rev B 1979; 19:4855.

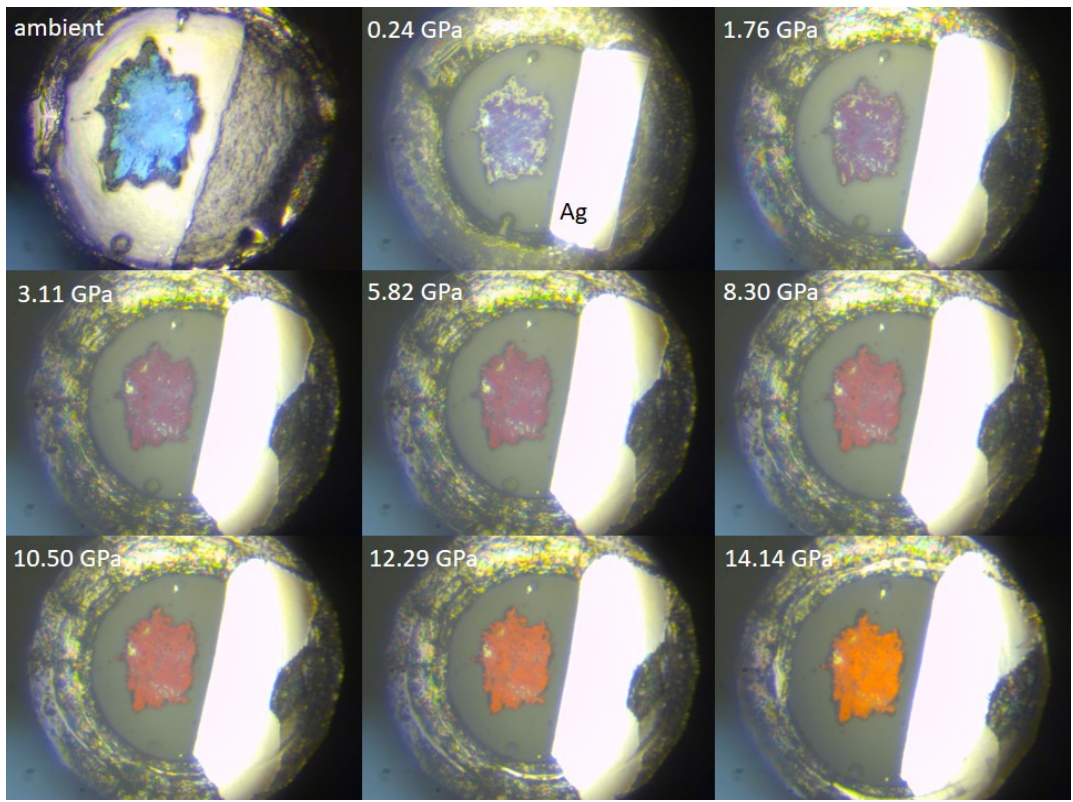

**Fig. 1.** Optical micrographs of the LuH$_2$ sample 1 at different pressures. A small piece of silver near the sample is used as the reflection reference. The color of the sample changes from shiny blue at ambient pressure to dark red around ~3 GPa, then becomes bright orange at higher pressures (>~12 GPa).

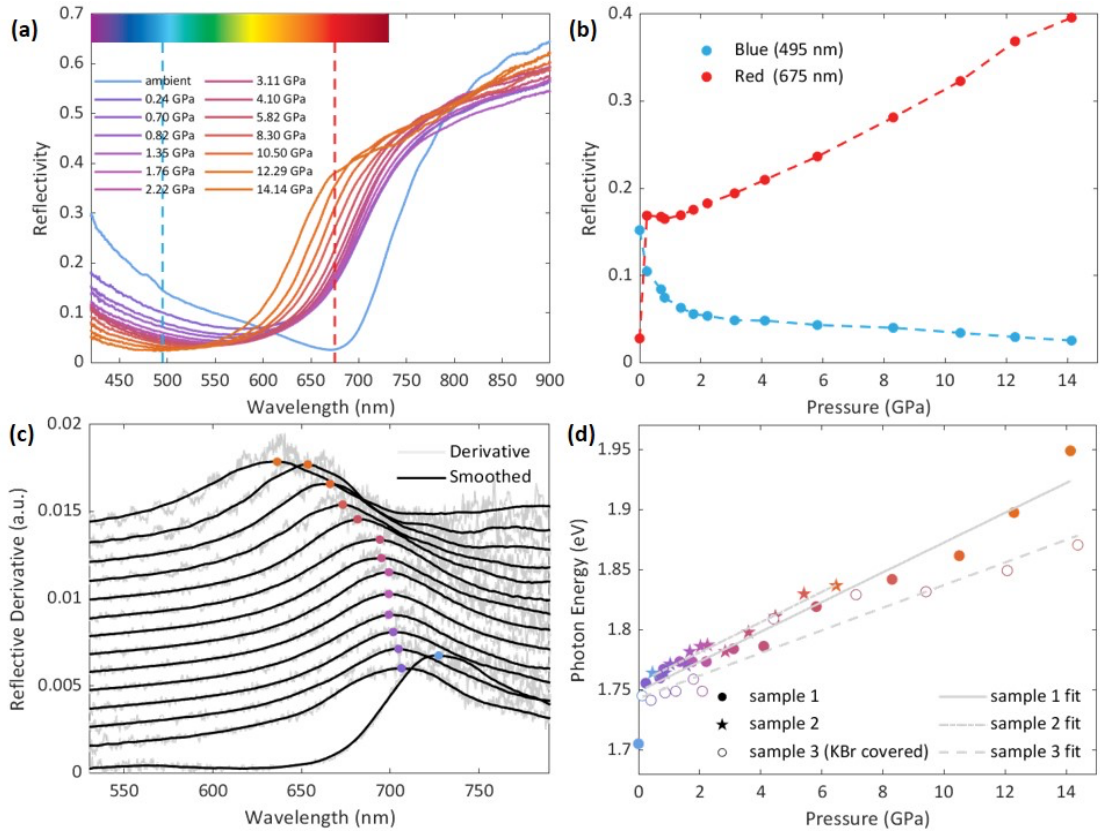

**Fig. 2.** (a) Optical reflectivity spectra of LuH$_2$ (sample 1) measured with varying pressures. The sharp change in reflectivity corresponds to the LuH$_2$ plasmon edge, which blueshifts when the pressure goes up, resulting in the increased reflection of red light and suppressed reflection of blue light. The process is consistent with the color change observed in Fig.1. (b) The evolution of reflectivity with pressure at two selective colors (495 nm blue and 675 nm red, respectively). (c) The plasmon resonance estimated from the peak of reflectivity derivative. (d) The plasmon energies as functions of pressure for sample 1-3 and the corresponding linear fits.

**Supplementary Materials**

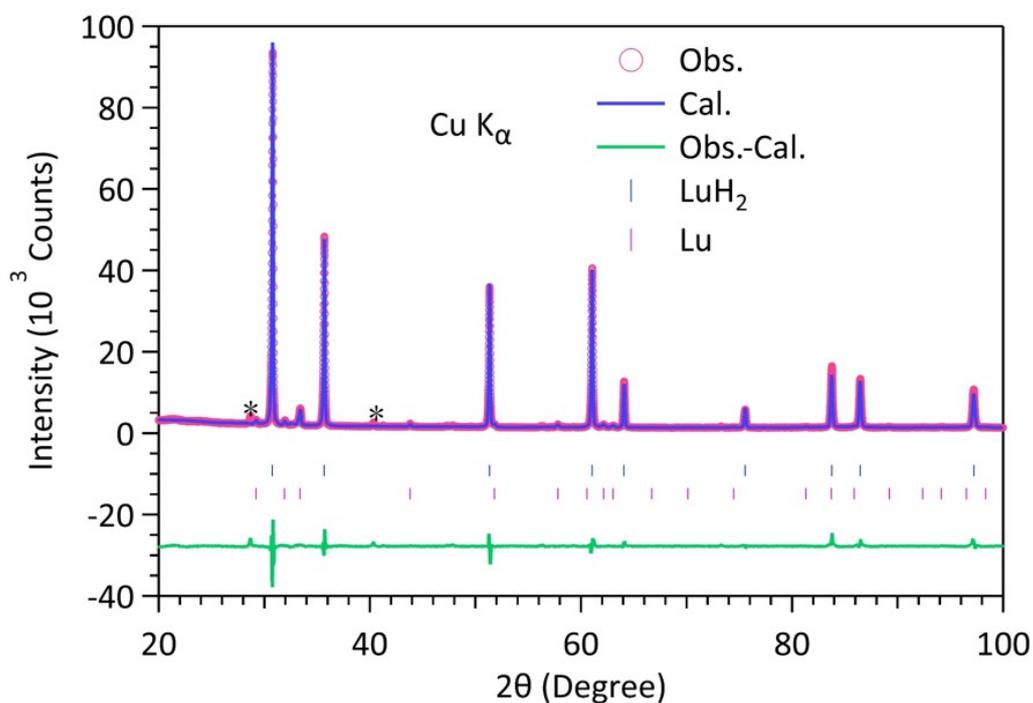

**Fig. S1.** Observed (open circle), calculated (solid line), and difference (bottom line) XRD profiles of the studied $LuH_2$ sample.

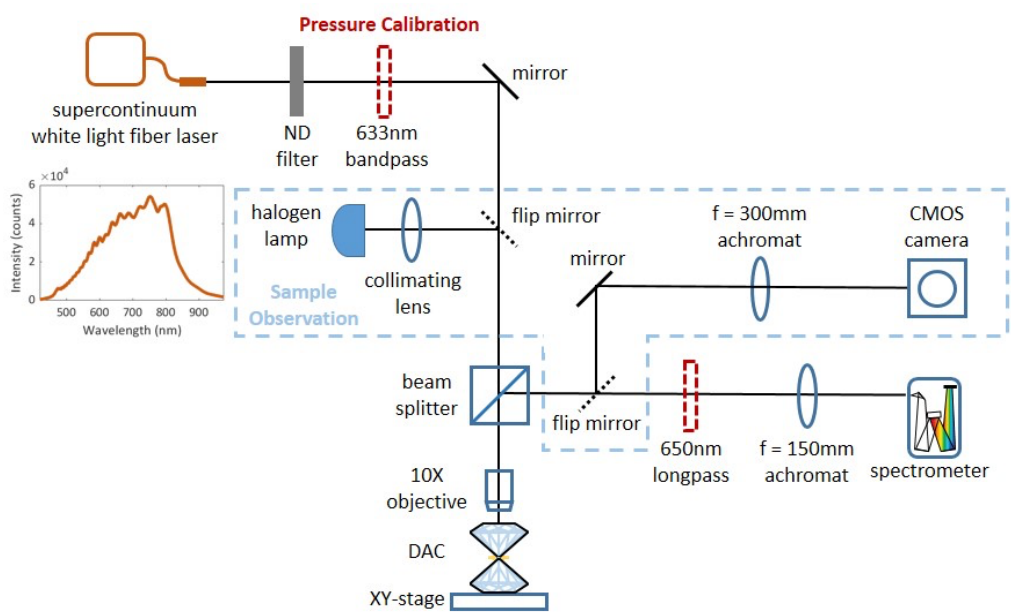

**Fig. S2.** Optical layout for sample observations, pressure calibrations and reflectance measurements (ND = Neutral Density). The plot below the supercontinuum laser shows the reference spectrum reflected from the silver mirror. A linear polarizer (not drawn) is used to minimize the potential spectrum distortion effects of the visible-range beamsplitter (Thorlabs, BS019). The laser beam is focused by a 10× objective (Motic, NA = 0.25, WD = 17.5). The DAC is placed on a XY linear stage to realize sample positioning.

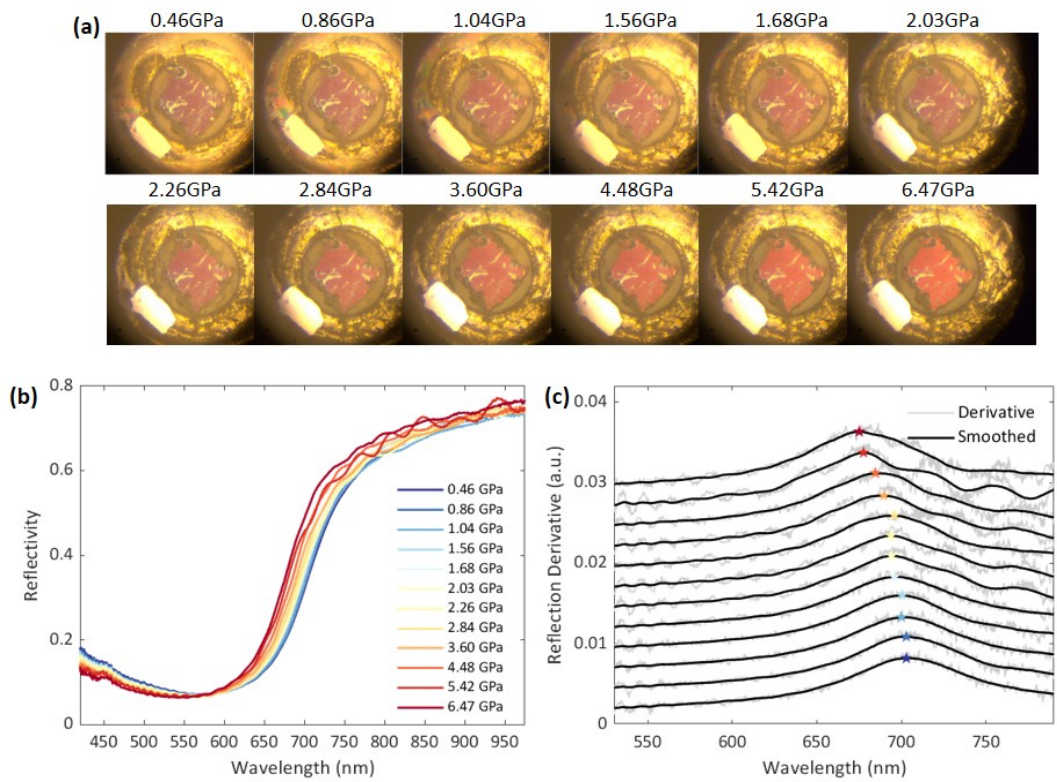

**Fig. S3.** Optical micrographs, reflectivity spectra, and data processing of sample 2 under different pressures.

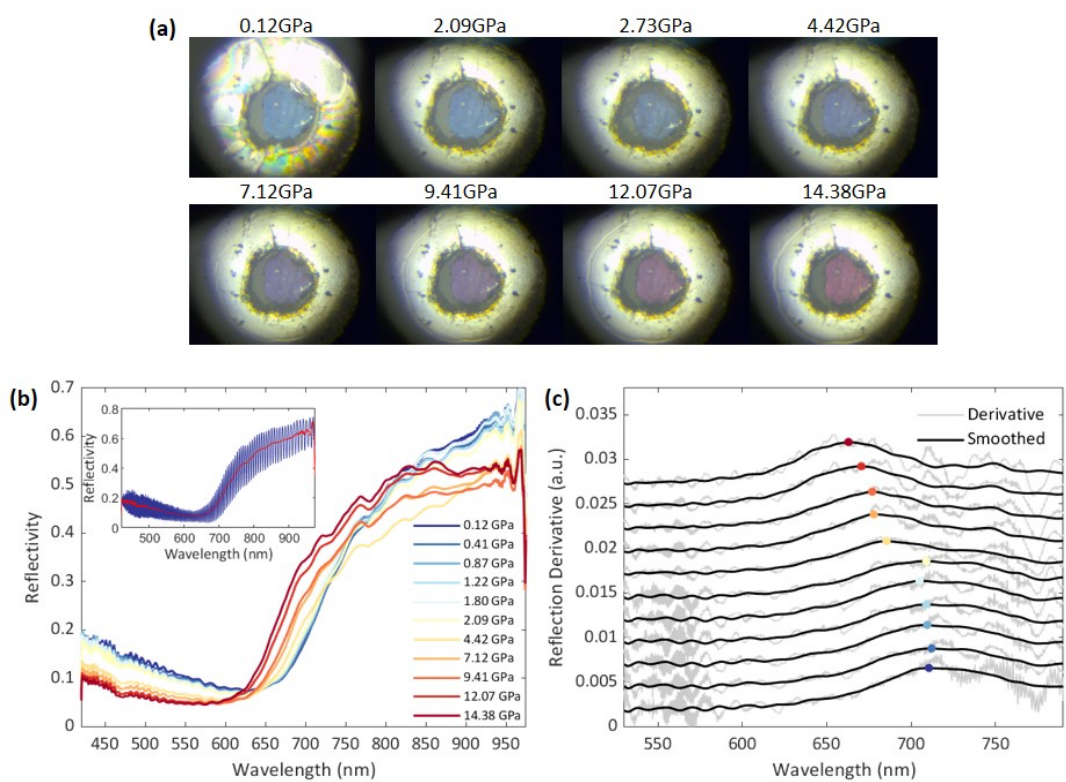

**Fig. S4.** Optical micrographs, reflectivity spectra, and data processing of sample 3 under different pressures (measured through KBr). At small pressures, the sample looks more bluish comparing to sample 1 and 2, where the high refractive index of the diamond plays a role in reducing the blue reflection of $LuH_2$ and modifying its general appearance to more violet-like. Inset of (b) displays an example of the raw data (indigo line) and FFT filtered data (red line) of the reflectivity spectrum at 0.12 GPa. Strong Fabry–Pérot interference emerges in the raw data due to the optical cavity formed between the upper surface of $LuH_2$ and the lower surface of the diamond anvil, yielding lower quality of the reflectivity data even after the FFT filtering. The extracted plasmon energy in c (also presented in Fig.2d) is less reliable comparing to sample 1 and 2.